\documentclass[11pt]{article}

\usepackage{graphicx} %插入图片的宏包
\usepackage{float} %设置图片浮动位置的宏包
\usepackage{subfigure} %插入多图时用子图显示的宏包
\usepackage{epstopdf}
\usepackage{color}
\usepackage{subfigure}
\usepackage{amsmath}
\usepackage{amssymb}
\usepackage{graphicx,color}
\usepackage{cite}
\usepackage{enumerate}
\usepackage{amsthm}% theorems and proofs
\usepackage{amsfonts,mathrsfs}
\usepackage{geometry}
\usepackage{ifthen}
\usepackage{color}
\usepackage{appendix}

\begin{document}

\title{\bf Quantumness of pure-state ensembles via coherence of Gram matrix based on generalized $\alpha$-$z$-relative R\'enyi entropy }

\vskip0.1in
\author{\small Wendao Yuan$^1$, Zhaoqi Wu$^1$\thanks{Corresponding author. E-mail: wuzhaoqi\_conquer@163.com}, Shao-Ming
Fei$^{2,3}$\\
{\small\it  1. Department of Mathematics, Nanchang University,
Nanchang 330031, P R China}\\
{\small\it  2. School of Mathematical Sciences, Capital Normal University, Beijing 100048, P R China}\\
{\small\it  3. Max-Planck-Institute for Mathematics in the Sciences,
04103 Leipzig, Germany} }
\date{}
\maketitle

\noindent {\bf Abstract} {\small } The Gram matrix of a set of
quantum pure states plays key roles in quantum information
theory. It has been highlighted that the Gram matrix of a pure-state
ensemble can be viewed as a quantum state, and the quantumness of a
pure-state ensemble can thus be quantified by the coherence of the
Gram matrix [Europhys. Lett. \textbf{134} 30003]. Instead of the $l_1$-norm
of coherence and the relative entropy of coherence, we utilize the generalized
$\alpha$-$z$-relative R\'enyi entropy of coherence of the Gram
matrix to quantify the quantumness of a
pure-state ensemble and explore its properties. We
show the usefulness of this quantifier by calculating the quantumness of six
important pure-state ensembles. Furthermore, we compare our quantumness with other existing ones and show their features as well as orderings.
\vskip 0.1 in

\vskip0.2in

\noindent {\bf 1. Introduction}

Defined by a finite set of vectors in an inner product space \cite{MA1985}, the Gram matrix has been extensively applied in many different branches of mathematics and physics. Notable features of the Gram matrix, including the eigenvalues \cite{FM2001}, the trace \cite{HK2018}, the determinant \cite{BT2010} and the entropy \cite{FN2012}, have been investigated. Recently, it has been shown that many important issues in quantum information theory, such as uncertainty relations \cite{RHP1934,GP2007,BKW2018}, state discrimination \cite{DD1988,FCA1996,JR2000,MA2007,PND2015,VEM2019}, transitions between two sets of quantum states \cite{CA2000,CA2004}, information-theoretic aspects of superposition \cite{CL2017}, quantum information masking \cite{qmasking} and PT-symmetric quantum systems \cite{WS2004}, are intimately related to the Gram matrix.

On the other hand, the characterization and quantification of the quantumness of ensembles have received extensive attention in the past few years. Various quantifiers of the quantumness of ensembles have been introduced \cite{LN2017,FS1998,SM2003,FCA2004,LSL2009,LSL2010,LSL2011,FCA2003,QX2018,LN2019,MY2019,HH2021}. Recently, it is pointed out by Sun, Luo and Lei \cite{SY2021} that the Gram matrix of a pure-state ensemble can be recognized as a quantum state. Based on this observation, a quantification of the quantumness of a pure-state ensemble has been proposed by exploiting the coherence of the Gram matrix of the ensemble.

Motivated by the work \cite{SY2021}, in this paper we adopt the
prior probability into the quantum state to form a Gram matrix, and
employ the generalized $\alpha$-$z$-relative R\'enyi entropy of
coherence of the Gram matrix to quantify the quantumness of the
corresponding pure-state ensemble. In Section 2, we review the
generalized $\alpha$-$z$-relative R\'enyi entropy and the related
coherence measure, and the Gram matrix of a pure-state ensemble and
its basic properties.  Then we provide the quantifier of the
quantumness of ensemble via coherence of the associated Gram matrix
in terms of the generalized $\alpha$-$z$-relative R\'enyi entropy.
We calculate this quantumness measure for six important ensembles
and compare with several other quantifiers of quantumness. We
conclude with a summary and some discussions in Section 3.

\vskip0.2in

\noindent {\bf 2. Quantumness of a pure-state ensemble via generalized $\alpha$-$z$-relative R\'enyi entropy of coherence}

Let $\mathcal{H}$ be a $d$-dimensional Hilbert space, and
$\mathcal{B(H)}$, $\mathcal{S(H)}$ and $\mathcal{D(H)}$ the set of
all bounded linear operators, Hermitian operators and density
operators on $\mathcal{H}$ (positive operators with trace 1),
respectively. Let $\{|i\rangle\}^d_{i=1}$ be an orthonormal basis of
$\mathcal{H}$. A state is called incoherent if the density matrix is
diagonal with respect to this basis. Denote by $\mathcal{I}$ the set
of all incoherent states, $\mathcal{I}=\{\delta\in
\mathcal{D(H)}|\delta=\sum_{i}p_i|i\rangle\langle i|,~p_i\geq
0,~\sum_{i}p_i=1\}$.

Let $\Phi$ be a completely positive trace preserving (CPTP) map,
$\Phi(\rho)=\sum_{i}K_i\rho K_i^\dag$,
where $K_i$ are Kraus operators satisfying $\sum_{i}K_i^\dag
K_i=I_{d}$ with $I_d$ the identity operator. $K_i$ are called
incoherent if $K_i^\dag \mathcal{I}K_i\in
\mathcal{I}$ for all $i$, and the map is called incoherent.
A well-defined coherence measure $C$ should satisfy the following
conditions \cite{BT2014}:
$(C_1)$ (Faithfulness) $C(\rho)\geq 0$ and $C(\rho)=0$ iff $\rho$ is
incoherent.
$(C_2)$ (Monotonicity) $C(\Phi(\rho))\leq C(\rho)$ for any
incoherent operation $\Phi$.
$(C_3)$ (Convexity) $C(\cdot)$ is a convex function of $\rho$, i.e.,
$\sum_{i}p_iC(\rho_i)\geq C(\sum_{i}p_i\rho_i)$,
where $p_i\geq 0 $ and $ \sum_{i}p_i=1$.
$(C_4)$ (Strong monotonicity) $C(\cdot)$ does not increase on
average under selective incoherent operations, i.e.,
$C(\rho)\geq \sum_{i}p_iC(\varrho_i)$,
where $p_i=\mathrm{Tr}(K_i\rho K_i^\dag)$ are probabilities and
$\varrho_i=\frac{K_i\rho K_i^\dag}{p_i}$ are the post-measurement states,
$K_i$ are incoherent Kraus operators.
The conditions $(C_3)$ and $(C_4)$ can be replaced equivalently by the following additivity coherence for
block-diagonal states \cite{YXD2016},
$C(p\rho_1\oplus(1-p)\rho_2)=pC(\rho_1)+(1-p)C(\rho_2)$.

\textcolor{blue}{Also, denote the support of an operator $\rho$ by
{supp} $\rho$. The support of an operator is defined to be the
vector space orthogonal to its kernel. For a Hermitian operator,
this means the vector space spanned by eigenvectors of the operator
with non-zero eigenvalues. For any two quantum states
$\rho,\sigma\in \mathcal{D(H)}$ with {supp} $\rho\subset$ {supp}
$\sigma$, the generalized $\alpha$-$z$-relative R\'enyi entropy is
defined by \cite{ZXN2019},
\begin{equation}\label{eq1}
D_{\alpha, z}(\rho , \sigma)=
\frac{f^{\frac{1}{\alpha}}_{\alpha,z}(\rho ,
\sigma)-1}{\alpha-1},~~~~\alpha\in (-\infty,1)\cup (1,+\infty),~z>0,
\end{equation}
where
\cite{XCH2018,AKMR2015,ZXN2019}
$$
f_{\alpha ,z}(\rho,\sigma)=\mathrm{Tr}(\sigma ^{\frac
{1-\alpha}{2z}}\rho ^{\frac{\alpha}{z}}\sigma
^{\frac{1-\alpha}{2z}})^z.
$$
Also, negative powers are defined in the sense of generalized inverses; that is, for negative $x$, $\rho^x:=(\rho|_{\rm{supp}\rho})^x$. }
For states $\rho$ and $\sigma$, (1) if $0<\alpha<1$ and $z>0$, we
have $f_{\alpha ,z}(\rho,\sigma)\leq1$; (2) if $\alpha>1$ and $z>0$,
we have $f_{\alpha ,z}(\rho,\sigma)\geq1$. \textcolor{blue}{It is
shown that when $\alpha\rightarrow1$ and $z=1$,
$D_{\alpha,z}(\rho,\sigma)$ reduces to
$S'(\rho||\sigma)=\mathrm{Tr}\rho\ln\rho-\mathrm{Tr}\rho\ln\sigma$,
where `ln' indicates a natural logarithm. Note that
$S'(\rho||\sigma)=\ln2\cdot S(\rho||\sigma)$, where
$S(\rho||\sigma)=\mathrm{Tr}\rho\log\rho-\mathrm{Tr}\rho\log\sigma$
is the standard relative entropy between two quantum states $\rho$
and $\sigma$, in which the logarithm `log' is taken to base $2$
\cite{ZH2018,RAE2016}.}

The quantum coherence $C_{\alpha ,z}(\rho)$ of a state $\rho$ is defined by \cite{ZXN2019},
\begin{equation}\label{eq2}
C_{\alpha ,z}(\rho) = \underset{\sigma\in
\mathcal{I}}{\min}D_{\alpha,z}(\rho,\sigma),
\end{equation}
which is a well-defined measure of coherence in the following cases \cite{ZXN2019}:
(i) $\alpha\in (0,1)$ and $z \ge \max\{\alpha , 1-\alpha \};$
(ii) $\alpha\in (1,2]$ and $z = 1;$
(iii) $\alpha\in (1,2]$ and $z = \frac{\alpha}{2}$;
(iv) $\alpha > 1 $ and $z = \alpha$. In particular, for $\alpha\in(0,1)\cup(1,2]$ and $z=1$, the generalized
$\alpha$-$z$-relative R\'enyi entropy of coherence can be written as \cite{ZXN2019},
\begin{equation}\label{eq3}
C_{\alpha,1}(\rho)=\frac{\sum_{i=1}^{d}\langle
i|\rho^{\alpha}|i\rangle^{\frac{1}{\alpha}}-1}{\alpha-1}.
\end{equation}
\textcolor{blue}{In a similar manner, when $\alpha\rightarrow1$,
$C_{\alpha,1}(\rho)$ reduces to $\ln2\cdot C_{rel}(\rho)$,
where $C_{rel}(\rho)$ denotes the relative entropy of coherence
defined in \cite{BT2014}.}

Instead of a set
$\mathcal{S}=\{|\psi_1\rangle,|\psi_2\rangle,\cdots,|\psi_n\rangle\}$
of $n$ pure states in $\mathcal{H}$, we consider a pure-state
ensemble, $\mathcal{E}=\{(p_i,|\psi_i\rangle)~:~i=1,2,\cdots,n\}$,
where $p_i>0$ and $\sum_i p_i=1$. With respect to the set of vectors
$\{\sqrt{p_1}|\psi_1\rangle,\sqrt{p_2}|\psi_2\rangle,\dots,\sqrt{p_n}|\psi_n\rangle\}$,
the Gram matrix of $\mathcal{E}$ is defined as \cite{SY2021},
\begin{equation}\label{eq4}
G_\mathcal{E}=(\sqrt{p_ip_j}\langle \psi_i|\psi_j\rangle ),
\end{equation}
which is an $n \times n$ matrix with elements $\sqrt{p_ip_j} \langle
\psi_i|\psi_j \rangle$. It is easy to see that the diagonal elements
of $G_\mathcal{E}$ are $p_i$ . It has been proved by Sun et al.
\cite{SY2021} that the Gram matrix of a pure-state ensemble
(\ref{eq4}) has the following properties.

(a) (State interpretation)
$G_\mathcal{E}$ is a non-negative semidefinite matrix satisfying
$\mathrm{Tr}G_{\mathcal{E}}=1$. $G_\mathcal{E}$ is diagonal if and
only if the pure states in the ensemble $\mathcal{E}$ are mutually
orthogonal.

(b) (Unitary invariance) $G_{U\mathcal{E}}=G_\mathcal{E}
$ for any unitary operator $U$ on $\mathcal{H}$, where $U\mathcal{E}
=\{(p_i, U|\psi _i \rangle ):i =1,2,\dots,n \}.$

(c) (Hadamard
multiplicability) Denote $\mathcal{E} _1\circ \mathcal{E} _2 =\{(p_i
q_i,|\psi_i \rangle \otimes|\phi_i\rangle): i=1,2,\dots,n\}$ for two
ordered quantum ensembles $\mathcal{E} _1=\{(p_i,|\psi_i \rangle ):
i = 1,2,\dots,n\}$ and $\mathcal{E} _2=\{(q_i,|\phi_i\rangle ): i =
1,2,\dots ,n\}$. Then $G_{\varepsilon_1\circ
\varepsilon_2}=G_{\varepsilon_1} \circ G_{\varepsilon_2}$, where
$A\circ B=(a_{ij}b_{ij})$ denotes the matrix Hadamard product of $n
\times n $ matrices $A=(a_{ij})$ and $B = (b_{ij})$.

(d) (Tensor
multiplicability) For any two quantum ensembles $\mathcal{E} =
\{(p_i,|\psi_i \rangle ):i=1,2,\dots ,n\}$ and
$\mathcal{F}=\{(q_k,|\phi_k \rangle):k=1,2,\dots,m\}$, denote
$\mathcal{E} \otimes \mathcal{F}=\{(p_iq_k,|\psi_i \rangle \otimes
|\phi _k \rangle ) : i = 1,2,\dots,n, k= 1,2,\dots,m\}$. Then
$G_{\mathcal{E} \otimes \mathcal{F}}=G_{\mathcal{E}} \otimes
G_{\mathcal{F}}$.

The cross Gram matrix between $\mathcal{E}$ and $\mathcal{F}$ is
defined by \cite{SY2021},
$G_{\mathcal{E},\mathcal{F}}=(\sqrt{p_iq_k}\langle\psi_i|\phi
_k\rangle)$. It has been proved that \cite{SY2021}
$G_{U\mathcal{E},U\mathcal{F}}=G_{\mathcal{E},\mathcal{F}}$ for any
unitary operator $U$ on $\mathcal{H}$. When
$\mathcal{E}=\mathcal{F}$, one has
$G_{\mathcal{E},\mathcal{E}}=G_{\mathcal{E}}$.

From the property (a), we can view $G_\mathcal{E}$ as a density matrix in an $n$-dimensional Hilbert space.
Let $\mathcal{E}=\{(p_i,|\psi_i \rangle):i = 1,2,\dots,n\}$ be a pure-state ensemble, and $G_{\mathcal{E}} = (\sqrt{p_ip_j}\langle
\psi_i |\psi_j \rangle)$ the corresponding Gram matrix. We define the quantumness of a pure-state ensemble $\mathcal{E}$ as the coherence of the Gram matrix $G_{\mathcal{E}}$ based on the generalized $\alpha$-$z$-relative R\'enyi entropy,
\begin{equation}\label{eq5}
Q_{\alpha,z}(\mathcal{E})=C_{\alpha,z}(G_\mathcal{E}).
\end{equation}
\textcolor{blue}{By Eqs. (\ref{eq1}) and (\ref{eq2}), Eq.
(\ref{eq5}) can be rewritten as,
\begin{equation}\label{eq6}
Q_{\alpha,z}(\mathcal{E})=\underset{\sigma\in \mathcal{I}}{\min}
\frac{f^{\frac{1}{\alpha}}_{\alpha,z}(G_\mathcal{E},
\sigma)-1}{\alpha-1}.
\end{equation}}

\textcolor{blue}{For any $\alpha,z$ satisfying one of the cases
(i)-(iv) below Eq. (2)}, the quantumness measure
$Q_{\alpha,z}(\cdot)$ has the following desirable properties.

(1) (Positivity) $Q_{\alpha,z}(\mathcal{E}) \ge 0$ with equality
holding if and only if $\mathcal{E}$ is a classical ensemble in the
sense that the pure states in the ensemble are mutually orthogonal.
This is due to that $C_{\alpha,z}(\cdot)$ is a well-defined coherence measure, namely, $C_{\alpha,z}(\mathcal{E})\ge 0$, which implies that
$Q_{\alpha ,z}(\mathcal{E}) \ge 0$. Moreover, $Q_{\alpha,z}(\mathcal{E})= 0$ iff $C_{\alpha,z}(G_{\mathcal{E}})= 0$ iff $G_{\mathcal{E}}$ is diagonal iff the pure states in the ensemble are pairwise orthogonal.

(2) (Unitary invariance) $Q_{\alpha,z}(\cdot)$ is unitary invariant
in the sense that $Q_{\alpha,z}(U\mathcal{E})=Q_{\alpha,z}(\mathcal{E})$ for any
unitary operator $U$ on $\mathcal{H}$, where $U\mathcal{E}= \{(p_i,
U|\psi_i\rangle): i=1,2,\dots,n\}$. This can be seen from the properties of the cross Gram matrix between two
pure-state ensembles. For any unitary operator $U$ on $\mathcal{H}$,
it holds that $G_{U\mathcal{E}}=G_{U\mathcal{E},U\mathcal{E}}=G_{\mathcal{E},\mathcal{E}}=G_{\mathcal{E}}$, which gives rise to
$Q_{\alpha ,z}(\mathcal{E})=Q_{\alpha ,z}(U\mathcal{E}).$

(3) (Subadditivity) $Q_{\alpha,z}(\cdot)$ is subadditive after normalization in the sense that
\begin{equation}\label{eq7}
Q_{\alpha,z}^\prime (\mathcal{E} \otimes \mathcal{F} )\le Q_{\alpha,z}^\prime (\mathcal{E}) + Q_{\alpha,z}^\prime (\mathcal{F}),
\end{equation}
for any two quantum ensembles $\mathcal{E} = \{(p_i,|\psi_i\rangle):
i = 1,2,\dots ,n \}$ and $\mathcal{F} = \{(q_k,|\phi_k\rangle): k =
1,2,\dots ,m \}$. \textcolor{blue}{Here $Q_{\alpha,z}^\prime
(\mathcal{E}) = Q_{\alpha,z}(\mathcal{E})/n$ and
$Q_{\alpha,z}^\prime (\mathcal{F})=Q_{\alpha,z}(\mathcal{F})/m$ with
$n$ and $m$ being the number of quantum states in the ensembles
$\mathcal{E}$ and $\mathcal{F}$, respectively}, and the tense
product of two quantum ensembles is defined as $\mathcal{E} \otimes
\mathcal{F} = \{(p_iq_k , |\psi_i\rangle\otimes |\phi_k\rangle):i=
1,2,\dots, n, k = 1,2,\dots, m \}.$ \textcolor{blue}{The proof of
property (3) is given in the appendix.}

We calculate the quantumness defined by (\ref{eq5}) for several important ensembles and compare them with other quantifiers of quantumness proposed in previous literatures.

\noindent {\bf \small Example 1}
Consider the B92 ensemble on $\mathbb{C}^2$ \cite{BCH1992},
\begin{equation*}
\mathcal{E}_x=\left \{\left(\frac{1}{2},|\psi_1\rangle\right),\left(\frac{1}{2},|\psi_2\rangle\right)\right\},
\end{equation*}
where $\langle\psi_1|\psi_2\rangle=\sin\theta=x,~\theta\in[0,\frac{\pi}{2}]$, and
$|\psi_1\rangle=\cos\frac{\theta}{2}|0\rangle+\sin\frac{\theta}{2}|1\rangle,~|\psi_2\rangle=\sin\frac{\theta}{2}|0\rangle+\cos\frac{\theta}{2}|1\rangle.$
The Gram matrix of $\mathcal{E}_x$ is
       %开始数学环境
$$G_{\mathcal{E}_x}=\frac{1}{2}\left(                 %左括号
  \begin{array}{cc}   %该矩阵一共2列，每一列都居中放置
    1 & x \\  %第一行元素
    x & 1\\  %第二行元素
  \end{array}
\right)     $$            % 右括号
with eigenvalues $\frac{1\pm x}{2}$. By direct computation we have the quantumness of $\mathcal{E}_x$,
\begin{equation}\label{ex}
Q_{\alpha,1}(\mathcal{E}_x)=\frac{2^{-\frac{1}{\alpha}}[(1-x)^\alpha+(1+x)^\alpha]^{\frac{1}{\alpha}}-1}{\alpha-1},
\end{equation}
which captures the overlap between $|\psi_1\rangle$ and $|\psi_2\rangle$. In particular, when $x=\frac{1}{\sqrt{2}}$, we have
\begin{equation}\label{b92}
Q_{\alpha,1}(\mathcal{E}_{\mathrm{B92}})=\frac{2^{-\frac{1}{\alpha}}[(1-\frac{1}{\sqrt{2}})^\alpha+(1+\frac{1}{\sqrt{2}})^\alpha]^{\frac{1}{\alpha}}-1}{\alpha-1},
\end{equation}
where $\mathcal{E}_{\mathrm{B92}}=\mathcal{E}_{\frac{1}{\sqrt{2}}}$,
\textcolor{blue}{and
$\lim_{\alpha\rightarrow1}Q_{\alpha,1}(\mathcal{E}_{\mathrm{B92}})\approx
0.28$.}

\noindent {\bf \small  Example 2} Consider the diagonal ensemble \cite{LSL2011}, $$\mathcal{E}_{\mathrm{diag}}=\left\{\left(\frac{1}{3},|0\rangle\right),\left(\frac{1}{3},|1\rangle\right),\left(\frac{1}{3},|+\rangle\right)\right\},$$
 where $|+\rangle=\frac{|0\rangle+|1\rangle}{\sqrt{2}}$. The Gram matrix of ${\mathcal{E}_{\mathrm{diag}}}$ is
 $$G_{\mathcal{E}_{\mathrm{diag}}}=\frac{1}{3}\left(                 % 左括号
  \begin{array}{ccc}   %该矩阵一共3列，每一列都居中放置
    1 & 0 & \frac{1}{\sqrt{2}} \\  %第一行元素
    0 & 1 & \frac{1}{\sqrt{2}} \\
    \frac{1}{\sqrt{2}} &\frac{1}{\sqrt{2}} & 1 \\
  \end{array}
\right)     $$
with eigenvalues $\frac{2}{3},\frac{1}{3},0$.  Direct computation shows that the quantumness of $\mathcal{E}_{\mathrm{diag}}$ is
\begin{equation}\label{diag}
Q_{\alpha,1}({\mathcal{E}_{\mathrm{diag}}})=\frac{2^{\frac{\alpha-1}{\alpha}}[(1+2^{\alpha-1})^{\frac{1}{\alpha}}+1]-3}{3(\alpha-1)},
\end{equation}
\textcolor{blue}{and
$\lim_{\alpha\rightarrow1}Q_{\alpha,1}(\mathcal{E}_{\mathrm{diag}})\approx
0.46$.}

\noindent {\bf \small  Example 3} Consider the trine ensemble \cite{PSJD2000,HAS1973,DE1978,PA1991,HP1994,BJC2005},
$$\mathcal{E}_{\mathrm{trine}}=\left\{\left(\frac{1}{3},|0\rangle\right),\left(\frac{1}{3},\frac{1}{2}|0\rangle+\frac{\sqrt{3}}{2}|1\rangle\right),\left(\frac{1}{3},\frac{1}{2}|0\rangle-\frac{\sqrt{3}}{2}|1\rangle\right)\right\}.$$
The Gram matrix of $\mathcal{E}_{\mathrm{trine}}$ is
$$G_{\mathcal{E}_{\mathrm{trine}}}=\frac{1}{6}\left(                 %左括号
  \begin{array}{ccc}   %该矩阵一共3列，每一列都居中放置
    2 & 1 & 1 \\  %第一行元素
    1 & 2 & -1 \\
    1 &-1 & 2 \\
  \end{array}
\right)     $$            % 右括号
with eigenvalues $\frac{1}{2},\frac{1}{2},0.$ Direct computation shows that the quantumness of $\mathcal{E}_{\mathrm{trine}}$ is
\begin{equation}\label{trine}
Q_{\alpha,1}({\mathcal{E}_{\mathrm{trine}}})=\frac{(\frac{2}{3})^{{\frac{1-\alpha}{\alpha}}}-1}{\alpha-1},
\end{equation}
\textcolor{blue}{and
$\lim_{\alpha\rightarrow1}Q_{\alpha,1}(\mathcal{E}_{\mathrm{trine}})\approx0.41$.}

\noindent {\bf \small  Example 4} Consider the BB84 ensemble \cite{BCH1984},
$$\mathcal{E}_{\mathrm{BB84}}=\left\{\left(\frac{1}{4},|0\rangle\right),\left(\frac{1}{4},|1\rangle\right),\left(\frac{1}{4},
|+\rangle\right),\left(\frac{1}{4},|-\rangle\right)\right\},$$
where $|-\rangle=\frac{|0\rangle-|1\rangle}{\sqrt{2}}$. The Gram matrix of $\mathcal{E}_{\mathrm{BB84}}$ is
$$G_{\mathcal{E}_{\mathrm{BB84}}}=\frac{1}{4\sqrt{2}}\left(                 %左括号
  \begin{array}{cccc}   % 该矩阵一共4列，每一列都居中放置
    \sqrt{2} & 0 & 1 & 1 \\  %第一行元素
    0 & \sqrt{2} & 1 & -1 \\
    1 & 1 & \sqrt{2} & 0 \\
    1 & -1 & 0 & \sqrt{2} \\
  \end{array}
\right)     $$            % 右括号
with eigenvalues $\frac{1}{2},\frac{1}{2},0,0$. Direct computation shows that the quantumness of $\mathcal{E}_{\mathrm{BB84}}$ is
\begin{equation}\label{bb84}
Q_{\alpha,1}({\mathcal{E}_{\mathrm{BB84}}})=\frac{2^{\frac{\alpha-1}{\alpha}}-1}{\alpha-1},
\end{equation}
\textcolor{blue}{and
$\lim_{\alpha\rightarrow1}Q_{\alpha,1}(\mathcal{E}_{\mathrm{BB84}})\approx0.69$.}

\noindent {\bf \small  Example 5} Consider the tetrad ensemble \cite{DE1978},
$$\mathcal{E}_{\mathrm{tetrad}}=\{(p_j,|\psi_j\rangle):j=1,2,3,4\}$$
with $p_j=\frac{1}{4}~(j=1,2,3,4)$, and
$$|\psi_1\rangle=|0\rangle,~~~|\psi_2\rangle=\frac{1}{\sqrt{3}}|0\rangle+\sqrt{\frac{2}{3}}|1\rangle,$$
$$|\psi_3\rangle=\frac{1}{\sqrt{3}}|0\rangle+e^{\frac{2\pi i}{3}}\sqrt{\frac{2}{3}}|1\rangle,~~~
|\psi_4\rangle=\frac{1}{\sqrt{3}}|0\rangle+e^{\frac{4\pi i}{3}}\sqrt{\frac{2}{3}}|1\rangle.$$
A symmetric informationally complete (SIC) set in a Hilbert space $\mathcal{H}$ with dimension $d$ \cite{RJM2004,SAJ2010} is a set of $d^2$ pure states $|\psi_j\rangle$ such that
$$|\langle\psi_j|\psi_k\rangle|^2=\frac{1}{d+1},~~~ j\neq k.$$
It is easy to see that $\{|\psi_j\rangle:j=1,2,3,4\}$ in  $\mathcal{E}_{\mathrm{tetrad}}$ is a SIC set in $\mathbb{C}^2$. The Gram matrix of the ensemble $\mathcal{E}_{\mathrm{tetrad}}$ is
$$G_{\mathcal{E}_{\mathrm{tetrad}}}=\frac{1}{4\sqrt{3}}\left(                 %左括号
  \begin{array}{cccc}   % 该矩阵一共4列，每一列都居中放置
    \sqrt{3} & 1 & 1 & 1 \\  %第一行元素
    1 & \sqrt{3} & i & -i \\
    1 & -i & \sqrt{3} & i \\
    1 & i & -i & \sqrt{3} \\
  \end{array}
\right)     $$            % 右括号
with eigenvalues  $\frac{1}{2},\frac{1}{2},0,0$. Direct computation shows that the quantumness of $\mathcal{E}_{\mathrm{tetrad}}$ is
\begin{equation}\label{tetrad}
Q_{\alpha,1}({\mathcal{E}_{\mathrm{tetrad}}})=\frac{2^{\frac{\alpha-1}{\alpha}}-1}{\alpha-1},
\end{equation}
\textcolor{blue}{and
$\lim_{\alpha\rightarrow1}Q_{\alpha,1}(\mathcal{E}_{\mathrm{tetrad}})\approx0.69$.}

\noindent {\bf \small  Example 6} Consider the six-state ensemble \cite{SAJ2010,BD1998,BPH1999,SZ2009,WKW1989}
$$\mathcal{E}_{\mathrm{six}}=\left\{\left(\frac{1}{6},|0_{\mu}\rangle\right),\left(\frac{1}{6},|1_{\mu}\rangle\right):\mu=x,y,z\right\}$$
on $\mathbb{C}^2$, where $|0_z\rangle=|0\rangle, ~|1_z\rangle=|1\rangle$, and
$$|0_x\rangle=\frac{|0\rangle+|1\rangle}{\sqrt{2}},~~~
|1_x\rangle=\frac{|0\rangle-|1\rangle}{\sqrt{2}},$$
$$|0_y\rangle=\frac{|0\rangle+i|1\rangle}{\sqrt{2}},~~~|1_y\rangle=\frac{|0\rangle-i|1\rangle}{\sqrt{2}}.$$
The Gram matrix of the ensemble $\mathcal{E}_{\mathrm{six}}$ is
$$G_{\mathcal{E}_{\mathrm{six}}}=\frac{1}{12}\left(                 %左括号
  \begin{array}{cccccc}   % 该矩阵一共4列，每一列都居中放置
    2 & 0 & 1+i & 1-i &  \sqrt{2}  &  \sqrt{2}\\  %第一行元素
    0 & 2 & 1-i & 1+i  &  \sqrt{2}  &  -\sqrt{2} \\
    1-i & 1+i & 2 & 0   &  \sqrt{2}  & -\sqrt{2}i \\
    1+i & 1-i & 0 & 2   &  \sqrt{2}  &  \sqrt{2}i  \\
    \sqrt{2} & \sqrt{2} & \sqrt{2} & \sqrt{2} & 2 & 0 \\
    \sqrt{2} & -\sqrt{2} & \sqrt{2}i & -\sqrt{2}i & 0 & 2 \\
  \end{array}
\right)     $$            % 右括号
with eigenvalues  $\frac{1}{2},\frac{1}{2},0,0,0,0$. Direct computation shows that the quantumness of $\mathcal{E}_{\mathrm{six}}$ is
\begin{equation}\label{six}
Q_{\alpha,1}({\mathcal{E}_{\mathrm{six}}})=\frac{3^{\frac{\alpha-1}{\alpha}}-1}{\alpha-1},
\end{equation}
\textcolor{blue}{and
$\lim_{\alpha\rightarrow1}Q_{\alpha,1}(\mathcal{E}_{\mathrm{six}})\approx1.10$.
The values of $\lim_{\alpha\rightarrow1}Q_{\alpha,1}(\cdot)$ for the
six pure-state ensembles in the above examples differ from the ones
$Q_{rel}(\cdot)$ in Table 1 of Ref. \cite{SY2021} by a constant
factor $\ln2$.}

\textcolor{blue}{Note that the BB84 ensemble, the six-state ensemble
and so on are all treated as complete mixed state, or classical
state in other words, if viewed from the entanglement. However, they
look differently from the quantumness defined via the generalized
$\alpha$-$z$-relative R\'enyi entropy of coherence.}

\begin{figure}[H] %H为当前位置，!htb为忽略美学标准，htbp 为浮动图形
\centering %图片居中
\includegraphics[width=0.7\textwidth]{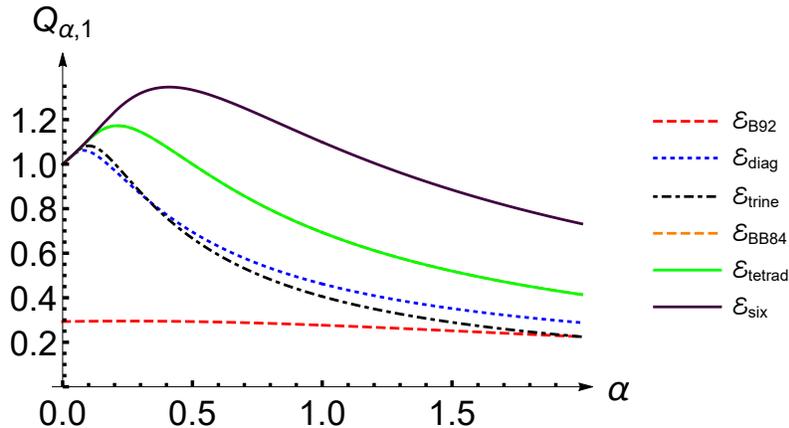} %插入图片，[]中设置图片大小，{}中是图片文件名
\caption{The quantumness $Q_{\alpha,1}(\cdot)$ of ensembles} %最终文档中希望显示的图片标题
\label{comp} %用于文内引用的标签
\end{figure}

It follows from Eqs. (\ref{bb84}) and (\ref{tetrad}) that
$Q_{\alpha,1}({\mathcal{E}_{\mathrm{BB84}}})=Q_{\alpha,1}({\mathcal{E}_{\mathrm{tetrad}}})$.
In addition, by Eqs. (\ref{b92}) and (\ref{trine}),
$Q_{2,1}({\mathcal{E}_{\mathrm{B92}}})=Q_{2,1}({\mathcal{E}_{\mathrm{trine}}})
=\sqrt{\frac{3}{2}}-1$.
We have the following observations, see  Figure \ref{comp}.
Among the six ensembles, $Q_{\alpha,1}({\mathcal{E}_{\mathrm{B92}}})$ is always the minimum, while $Q_{\alpha,1}({\mathcal{E}_{\mathrm{six}}})$ is always the maximum for all $\alpha$. For any fixed $\alpha$, one has the following ordering,
$$Q_{\alpha,1}(\mathcal{E}_{\mathrm{B92}})\le Q_{\alpha,1}(\mathcal{E}_{\mathrm{trine}})\le Q_{\alpha,1}(\mathcal{E}_{\mathrm{BB84}})= Q_{\alpha,1}(\mathcal{E}_{\mathrm{tetrad}})\le Q_{\alpha,1}(\mathcal{E}_{\mathrm{six}})$$
and
$$Q_{\alpha,1}(\mathcal{E}_{\mathrm{B92}})\le Q_{\alpha,1}(\mathcal{E}_{\mathrm{diag}})\le Q_{\alpha,1}(\mathcal{E}_{\mathrm{BB84}})= Q_{\alpha,1}(\mathcal{E}_{\mathrm{tetrad}})\le Q_{\alpha,1}(\mathcal{E}_{\mathrm{six}}).$$
The curves of $Q_{\alpha,1}(\mathcal{E}_{\mathrm{BB84}})$ and $ Q_{\alpha,1}(\mathcal{E}_{\mathrm{tetrad}})$ coincides as $Q_{\alpha,1}({\mathcal{E}_{\mathrm{BB84}}})=Q_{\alpha,1}({\mathcal{E}_{\mathrm{tetrad}}})$.
There is no ordering between $Q_{\alpha,1}(\mathcal{E}_{\mathrm{trine}})$ and $Q_{\alpha,1}(\mathcal{E}_{\mathrm{diag}})$ for $\alpha\in(0,1)\cup(1,2]$ in general.
In fact, $Q_{\alpha,1}(\mathcal{E}_{\mathrm{trine}})= Q_{\alpha,1}(\mathcal{E}_{\mathrm{diag}})$ when $\alpha=\alpha_*\approx0.33$, and we have $Q_{\alpha,1}(\mathcal{E}_{\mathrm{trine}}) \geq Q_{\alpha,1}(\mathcal{E}_{\mathrm{diag}}) $ when $\alpha \in(0,\alpha_*)$, while $Q_{\alpha,1}(\mathcal{E}_{\mathrm{trine}}) \le Q_{\alpha,1}(\mathcal{E}_{\mathrm{diag}}) $ when $\alpha\in(\alpha_*,1)\cup(1,2]$.

In order to get a more intuitive picture of the quantumness of quantum ensembles, we next compare our quantifiers with some existing ones in the literatures. For a ensemble $\mathcal{E}=\{(p_i,|\psi_i\rangle):i=1,2,\dots,n\}$ on a Hilbert space $\mathcal{H}$, the quantumness based on the $l_1$-norm of coherence of the Gram matrix is defined by \cite{BT2014,SA2017,SLH2014,RS2016},
$$Q_{l_1}(\mathcal{E})=\sum_{i\neq j}\sqrt{p_ip_j}|\langle\psi_i|\psi_j\rangle|.$$
The quantumness based on the security of information transmission is defined by \cite{FS1998,SM2003,LSL2011,FCA2003},
$$Q_{\mathrm{FS}}(\mathcal{E})=1-\sup_{M,\{\sigma_k\}}\sum_{i,k}p_i\mathrm{Tr}(|\psi_i\rangle\langle\psi_i|M_k)\mathrm{Tr}(|\psi_i\rangle\langle\psi_i|\sigma_k),$$
where the sup carries out with respect to all measurements $M=\{M_k\}$ on $\mathcal{H}$ and sets of quantum states $\{\sigma_k\}$. In \cite{LSL2011}
the quantumness based on quantum cloning is defined to be,
$$Q_{\mathrm{clon}}(\mathcal{E})=1-\sup_{U}\sum_ip_i|\langle\psi_i|\otimes\langle\psi_i|U|\psi_i\rangle\otimes|0\rangle|^2,$$
where the sup goes over all unitary operators $U$ on the composite system $\mathcal{H}\otimes\mathcal{H}$ such that $U(|\psi_i\rangle\otimes|0\rangle)=|\Psi_i\rangle$ has the same marginals, and $|0\rangle\in\mathcal{H}$ is any fixed pure state. Instead of $Q_{\mathrm{clon}}(\mathcal{E})$, a modified version $Q_{\mathrm{clon}}^{'}(\mathcal{E})$ is also considered by implementing a symmetric unitary operator when optimizing over $U$, which is easier to calculate.
The quantumness based on the Holevo quantity and the accessible information is given by \cite{LSL2011,HAS1973},
$$Q_{\mathrm{Hol}}(\mathcal{E})=\chi(\mathcal{E})-\chi_0(\mathcal{E}),$$
where
\begin{equation*}
\chi(\mathcal{E})=S(\sum_ip_i|\psi_i\rangle\langle\psi_i|)-\sum_ip_iS(|\psi_i\rangle\langle\psi_i|)
\end{equation*}%$$\chi(\mathcal{E})=S(\sum_ip_i|\psi_i\rangle\langle\psi_i|).$$
is the Holevo quantity of the pure-state ensemble $\mathcal{E}=\{(p_i,|\psi_i\rangle):i=1,2,\cdots,n\}$, while $\chi_0(\mathcal{E})=\sup_MI(M(\mathcal{E}))$ is the accessible information, in which the sup is conducted over all measurements $M=\{M_k$\} on $\mathcal{H}$, and
$$I(M(\mathcal{E}))=-\sum_ip_i\log p_i-\sum_k q_k\log q_k+\sum_{ik}q_{ik}\log q_{ik}$$
denotes the mutual information of the joint probability
distribution $q_{ik}=p_i\mathrm{Tr}(|\psi_i\rangle\langle\psi_i|M_k)$
with marginals $\{p_i=\sum_kq_{ik}\}$ and $\{q_k=\sum_iq_{ik}\}$.
%The Holevo inequality states that $\chi_0(\mathcal{E}\le\chi(\mathcal{E}))$, and the equality holds if and only if the states in the ensemble are mutually orthogonal\cite{HAS1973}.
The quantumness based on the commutator is defined by \cite{LN2017,QX2018,LN2019},
$$Q(\mathcal{E})=-\sum_{i,j}\sqrt{p_ip_j}\mathrm{Tr}[|\psi_i\rangle\langle\psi_i|,|\psi_j\rangle\langle\psi_j|]^2,$$
and
$$Q_{\mathrm{comm}}(\mathcal{E})=-\sum_{i,j}p_ip_j\mathrm{Tr}[|\psi_i\rangle\langle\psi_i|,|\psi_j\rangle\langle\psi_j|]^2.$$

Combining Eqs. (\ref{b92})-(\ref{six}), the Table 2 in \cite{SY2021} and the results in \cite{LSL2011}, we have the Table \ref{tab:comparing}, which gives a comparision among the different quantifiers of quantumness for these pure-state ensembles,
 \begin{table}[h]
        \centering
\setlength{\abovecaptionskip}{0pt}%
\setlength{\belowcaptionskip}{10pt}%
        \caption{Comparision of different quantifiers of quantumness of pure-state ensembles}
        \begin{tabular}{|l|c|c|c|c|c|c|}\hline
             &$\mathcal{E}_{\mathrm{B92}}$ &$\mathcal{E}_{\mathrm{diag}}$&$\mathcal{E}_{\mathrm{trine}}$&$\mathcal{E}_{\mathrm{BB84}}$ &$\mathcal{E}_{\mathrm{tetrad}}$&$\mathcal{E}_{\mathrm{six}}$\\\hline
            %&mm&inches\\
            $Q_{\alpha,1}$&Eq.(\ref{b92})&Eq.(\ref{diag})&Eq.(\ref{trine})&Eq.(\ref{bb84})&Eq.(\ref{tetrad})&Eq.(\ref{six})\\\hline
            $Q_{l_1}$&0.71&0.94&1&1.41&1.73&2.83\\\hline        $Q_{\mathrm{FS}}$&0.07&0.13&0.25&0.25&0.33&0.33\\\hline         $Q_{\mathrm{clon}}^{'}$&0.02&0.10&0.32&0.32&0.34&0.35\\\hline
            $Q_{\mathrm{Hol}}$&0.20&0.25&0.42&0.50&0.59&0.67\\\hline
            $Q_{\mathrm{comm}}$&0.25&0.22&0.25&0.25&0.33&0.33\\\hline
            $Q$&0.50&0.67&0.75&1&1.33&2\\\hline
        \end{tabular}

        \label{tab:comparing}
    \end{table}

In Figure 2 we plot the quantumness of the six ensembles based on different quantifiers. We have the following observations:

(1) For $Q_{\mathrm{comm}}(\cdot)$, the quantumness of the ensemble
$\mathcal{E}_{\mathrm{diag}}$ is the minimum, while the quantumness
of ensemble $\mathcal{E}_{\mathrm{six}}$ is the maximum. In
comparison, for other quantifiers, the quantumness of
$\mathcal{E}_{\mathrm{B92}}$ is the minimum, while the quantumness
of $\mathcal{E}_{\mathrm{six}}$ remains the maximum. For any fixed
$\alpha$, $Q_{\mathrm{comm}}(\cdot)$ yields the following ordering
for quantumness of ensembles,
$$\mathcal{E}_{\mathrm{diag}}\preceq\mathcal{E}_{\mathrm{B92}}\preceq \mathcal{E}_{\mathrm{trine}}\preceq \mathcal{E}_{\mathrm{BB84}}\preceq  \mathcal{E}_{\mathrm{tetrad}}\preceq \mathcal{E}_{\mathrm{six}},$$
while other quantifiers yield consistent orderings for quantumness of ensembles,
$$\mathcal{E}_{\mathrm{B92}}\preceq \mathcal{E}_{\mathrm{diag}}\preceq \mathcal{E}_{\mathrm{BB84}}\preceq  \mathcal{E}_{\mathrm{tetrad}}\preceq \mathcal{E}_{\mathrm{six}}$$
and
$$\mathcal{E}_{\mathrm{B92}}\preceq \mathcal{E}_{\mathrm{trine}}\preceq \mathcal{E}_{\mathrm{BB84}}\preceq  \mathcal{E}_{\mathrm{tetrad}}\preceq \mathcal{E}_{\mathrm{six}}.$$

(2) Some quantifiers yield strict orderings for the considered ensembles mentioned in observation (1), while other ones yield the same values for two or three ensembles, as pointed out in Ref. \cite{SY2021}.

(3) For a given ensemble among the considered ones, it is shown that $Q_{\alpha,1}(\cdot)$ coincides with the quantumness based on one or more quantifiers for different $\alpha$.
\begin{itemize}
\item[$\bullet$] For the B92 ensemble, the quantumness $Q_{\alpha,1}(\mathcal{E}_{\mathrm{B92}})$ = $Q_{\mathrm{comm}}(\mathcal{E}_{\mathrm{B92}})$ when $\alpha\approx1.53$.
\item[$\bullet$] For the diagonal ensemble, the quantumness $Q_{\alpha,1}(\mathcal{E}_{\mathrm{diag}})$= $Q_{l_1}(\mathcal{E}_{\mathrm{diag}})$ when $\alpha\approx0.23$; the quantumness $Q_{\alpha,1}(\mathcal{E}_{\mathrm{diag}})$= $Q(\mathcal{E}_{\mathrm{diag}})$ when $\alpha\approx0.54$.
\item[$\bullet$] For the trine ensemble, the quantumness $Q_{\alpha,1}(\mathcal{E}_{\mathrm{trine}})$ = $Q_{l_1}(\mathcal{E}_{\mathrm{trine}})$ when $\alpha\approx0.20$; the quantumness $Q_{\alpha,1}(\mathcal{E}_{\mathrm{trine}})$ = $Q_{\mathrm{FS}}(\mathcal{E}_{\mathrm{trine}})$ and $Q_{\alpha,1}(\mathcal{E}_{\mathrm{trine}})$= $Q_{\mathrm{comm}}(\mathcal{E}_{\mathrm{trine}})$ when $\alpha\approx1.77$; the quantumness $Q_{\alpha,1}(\mathcal{E}_{\mathrm{trine}})$= $Q_{\mathrm{clon}}^{'}(\mathcal{E}_{\mathrm{trine}})$ when $\alpha\approx1.33$; the quantumness $Q_{\alpha,1}(\mathcal{E}_{\mathrm{trine}})$ = $Q_{\mathrm{Hol}}(\mathcal{E}_{\mathrm{trine}})$ when $\alpha\approx0.96$; the quantumness $Q_{\alpha,1}(\mathcal{E}_{\mathrm{trine}})$ =$Q(\mathcal{E}_{\mathrm{trine}})$ when $\alpha\approx0.41$.
\item[$\bullet$] For the BB84 ensemble, the quantumness $Q_{\alpha,1}(\mathcal{E}_{\mathrm{BB844}})$= $Q_{\mathrm{Hol}}(\mathcal{E}_{\mathrm{BB844}})$ when $\alpha\approx1.59$; the quantumness $Q_{\alpha,1}(\mathcal{E}_{\mathrm{BB844}})$=$Q(\mathcal{E}_{\mathrm{BB844}})$ when $\alpha=0.50$.
\item[$\bullet$] For the tetrad ensemble, the quantumness $Q_{\alpha,1}(\mathcal{E}_{\mathrm{tetrad}})$ = $Q_{\mathrm{Hol}}(\mathcal{E}_{\mathrm{tetrad}})$ when $\alpha\approx1.26$.
\end{itemize}
%Comparing the Quantumness of various ensembles via coherence of Gram matrix based on the generalized $\alpha$-$z$-relative R\'enyi entropy of coherence.
\begin{figure}[H]
\centering
\subfigure[]%The quantumness of the B92 ensemble based on Gram matrix]
{
\begin{minipage}[t]{0.45\linewidth}
\centering
\includegraphics[width=1\textwidth,height=0.7\textwidth]{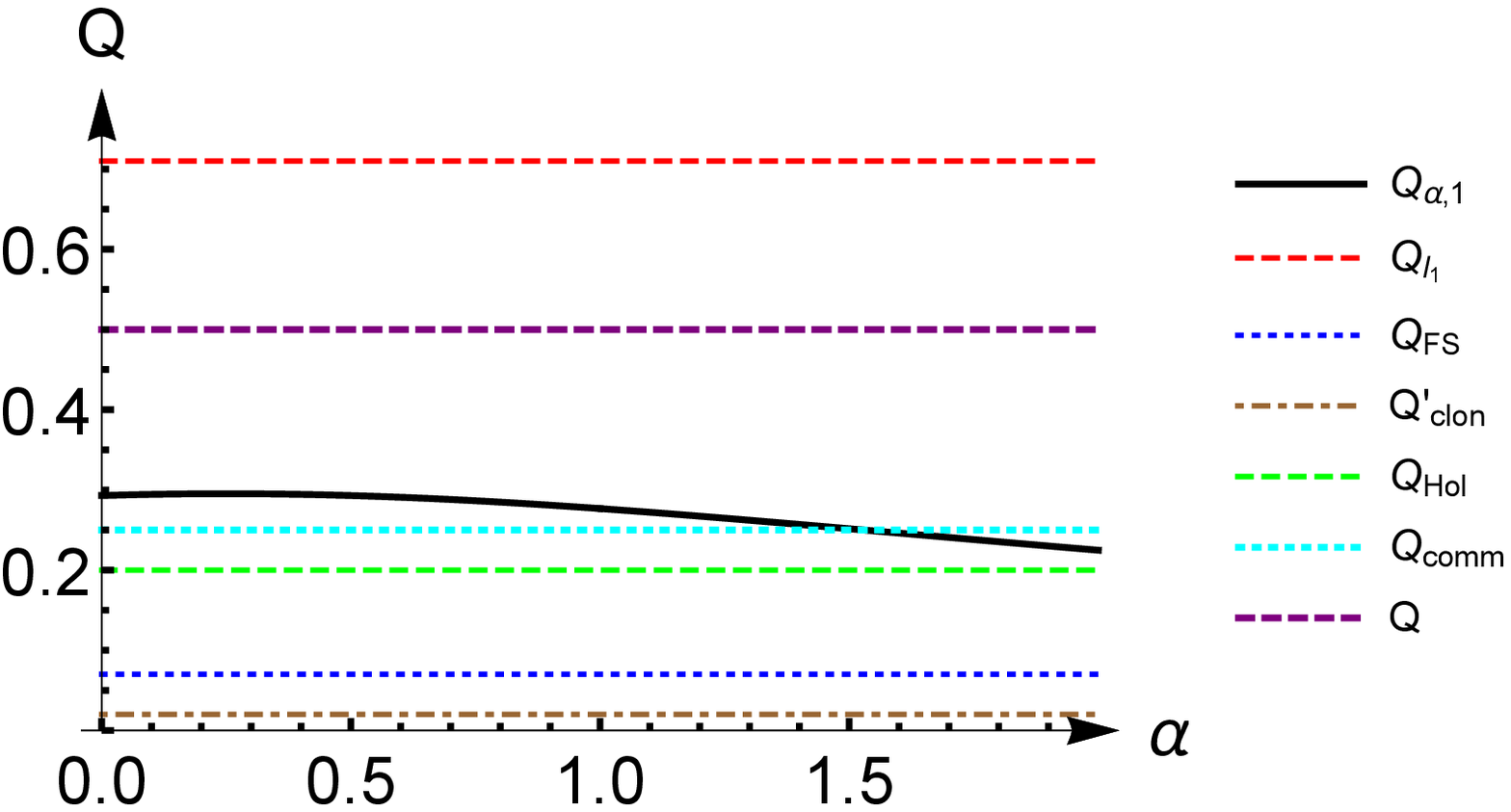}
%\caption{The quantumness of the B92 ensemble based on Gram matrix}
\end{minipage}%
}%\quad
\quad
\subfigure[]%The quantumness of the trine ensemble based on Gram matrix]
{
\begin{minipage}[t]{0.45\linewidth}
\centering
\includegraphics[width=1\textwidth,height=0.7\textwidth]{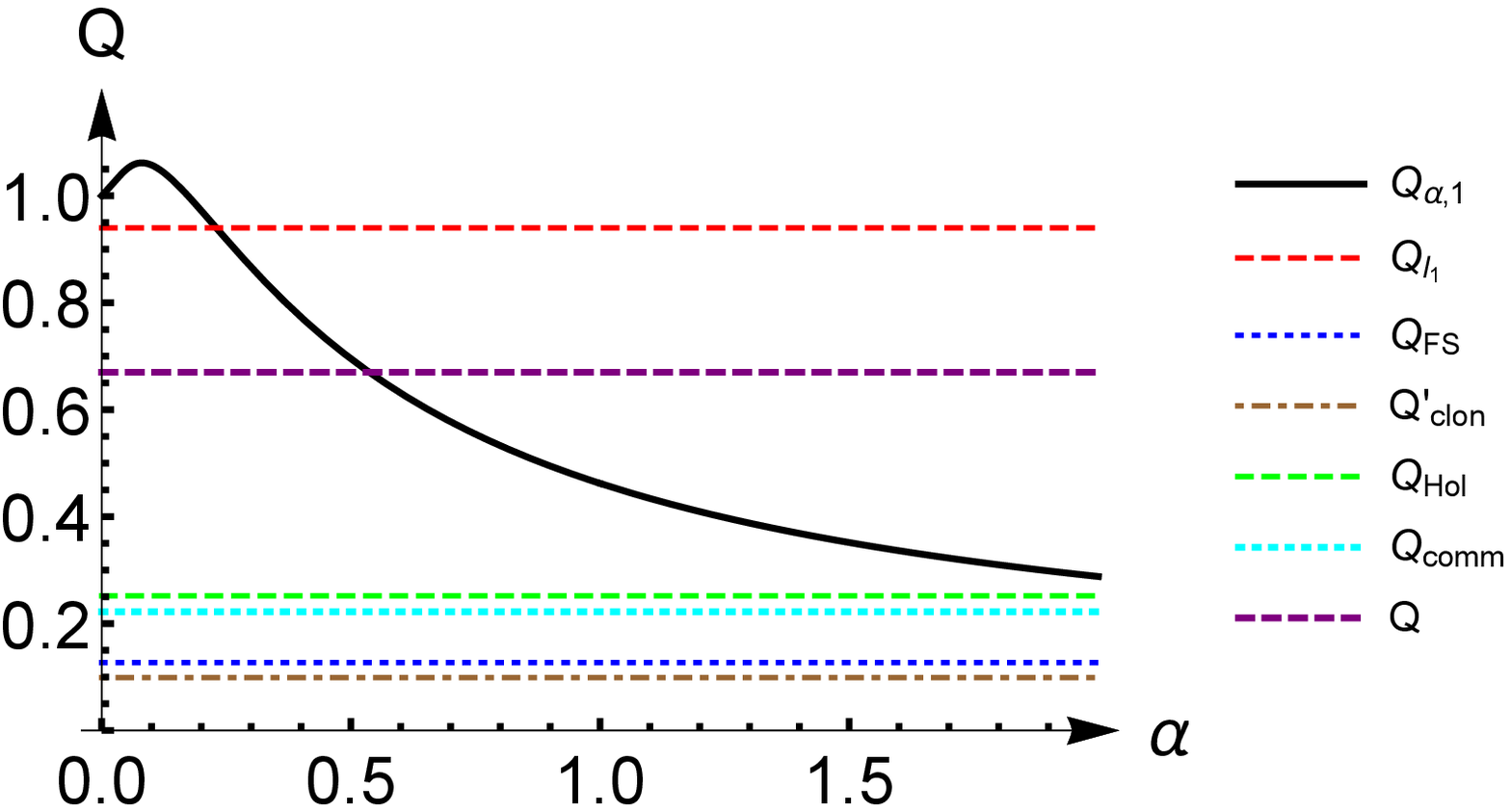}
%\caption{fig2}
\end{minipage}%
}%
\centering
%\caption{ pics}
\end{figure}
\begin{figure}[H]
\centering
\subfigure[]%The quantumness of the diagonal ensemble based on Gram matrix]
{
\begin{minipage}[t]{0.45\linewidth}
\centering
\includegraphics[width=1\textwidth,height=0.7\textwidth]{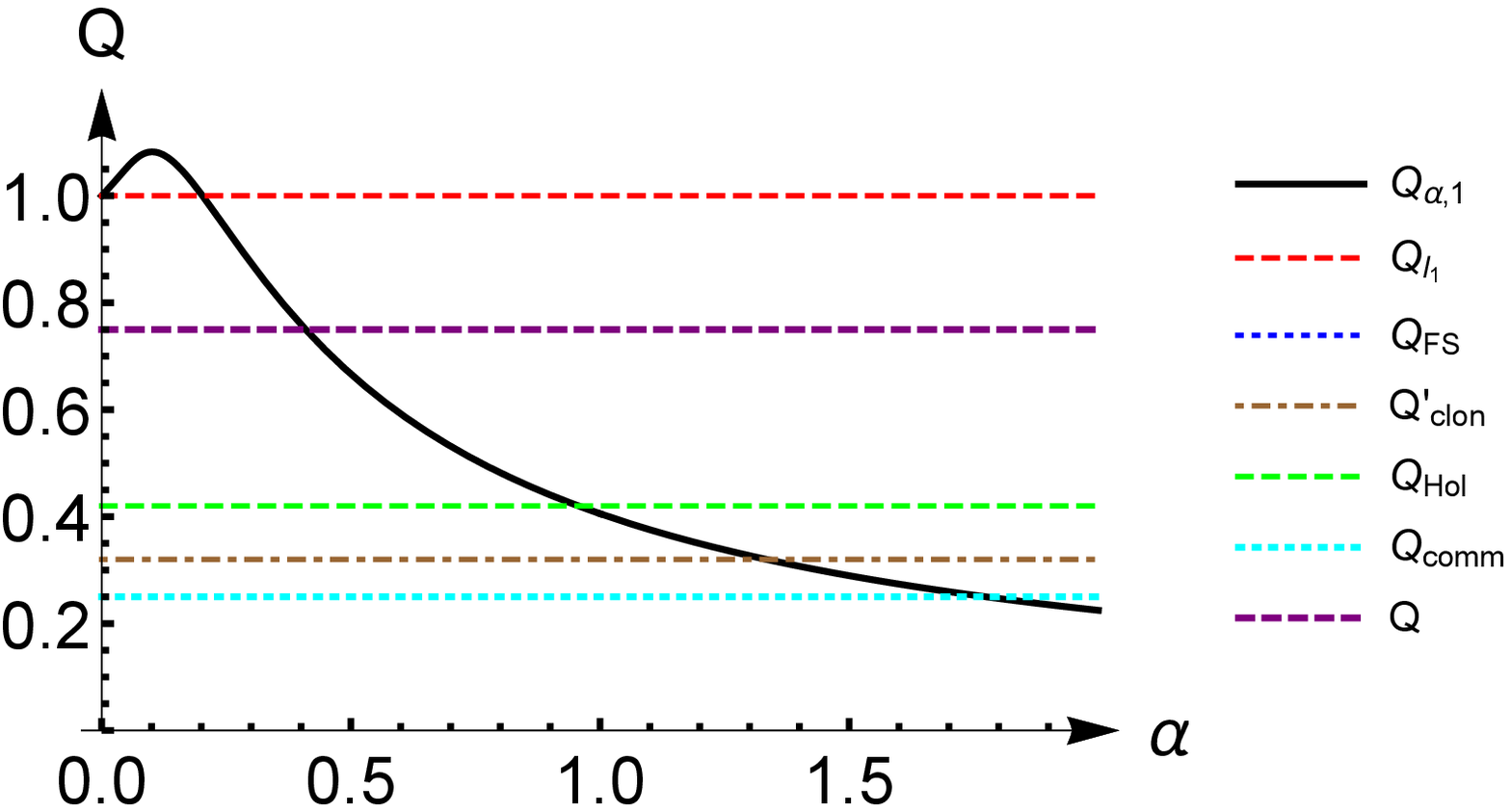}
%\caption{The quantumness of the B92 ensemble based on Gram matrix}
\end{minipage}%
}%\quad
\quad
\subfigure[]%The quantumness of the BB84 ensemble based on Gram matrix]
{
\begin{minipage}[t]{0.45\linewidth}
\centering
\includegraphics[width=1\textwidth,height=0.7\textwidth]{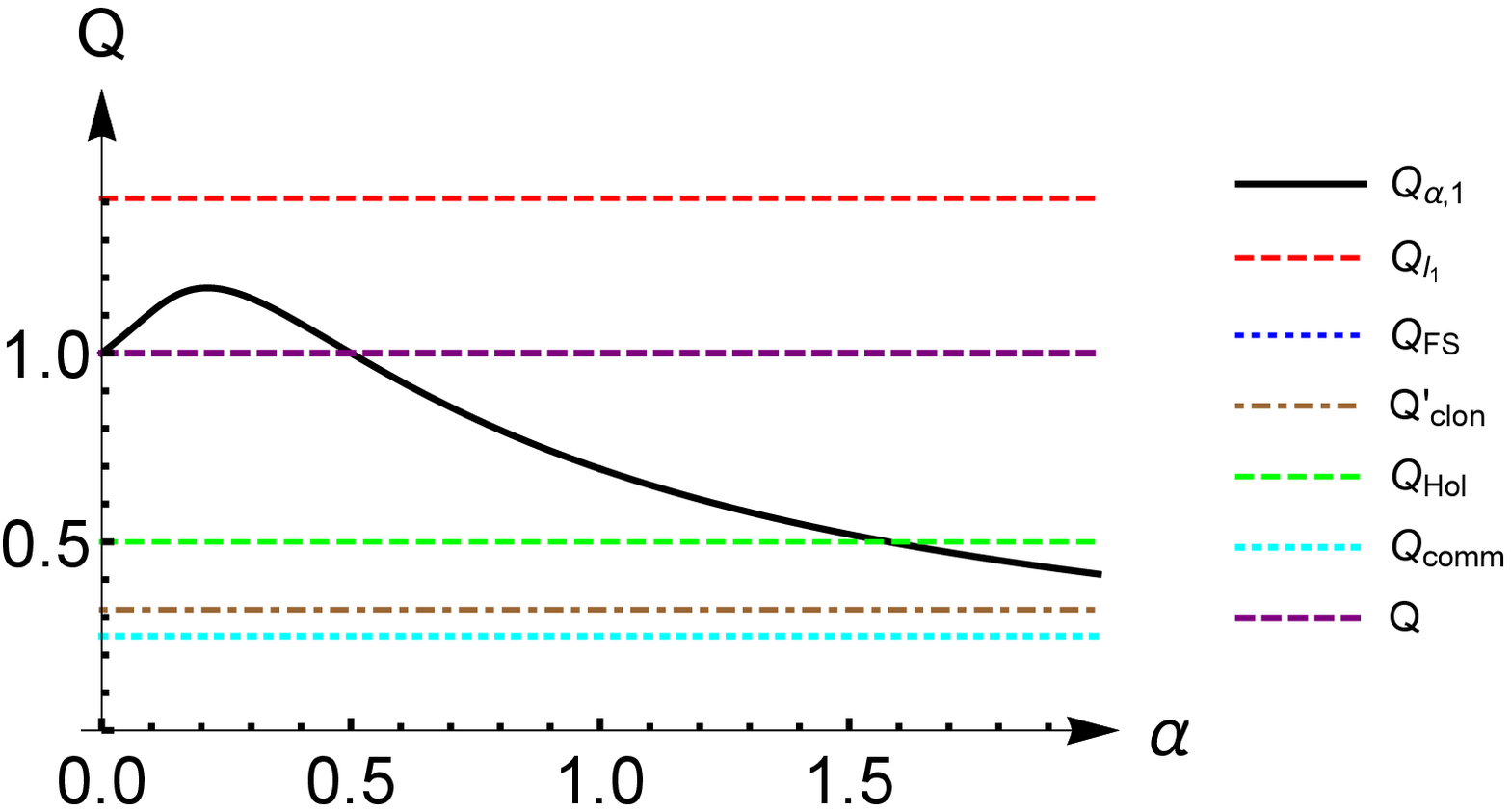}
%\caption{fig2}
\end{minipage}%
}%
\centering
%\caption{ pics}
\end{figure}

\begin{figure}[H]
\centering
\subfigure[]%The quantumness of the tetrad ensemble based on Gram matrix]
{
\begin{minipage}[t]{0.45\linewidth}
\centering
\includegraphics[width=1\textwidth,height=0.7\textwidth]{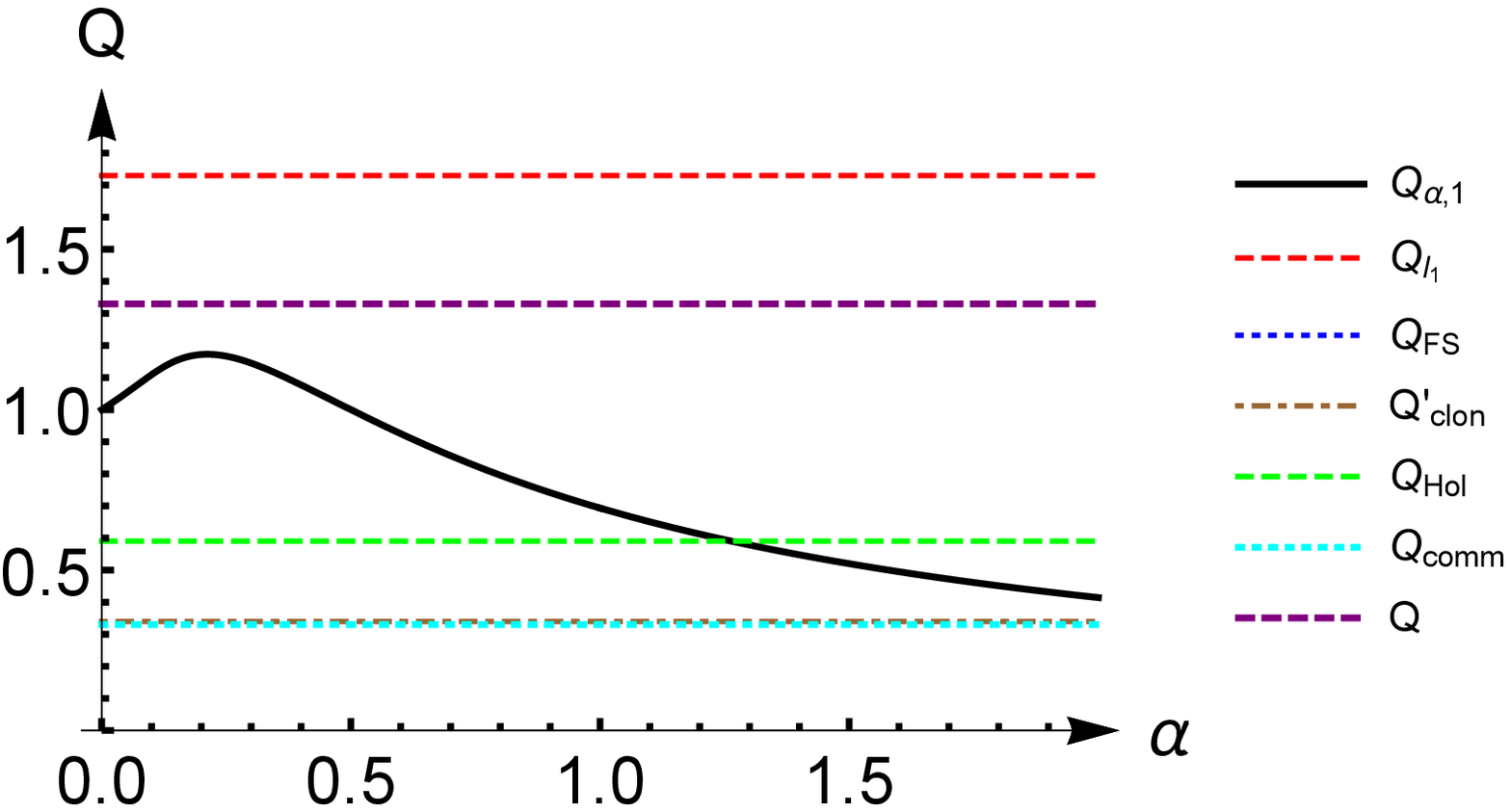}
%\caption{fig1}
\end{minipage}%
}%
\quad
\subfigure[]%The quantumness of the six-state ensemble based on Gram matrix]
{
\begin{minipage}[t]{0.45\linewidth}
\centering
\includegraphics[width=1\textwidth,height=0.7\textwidth]{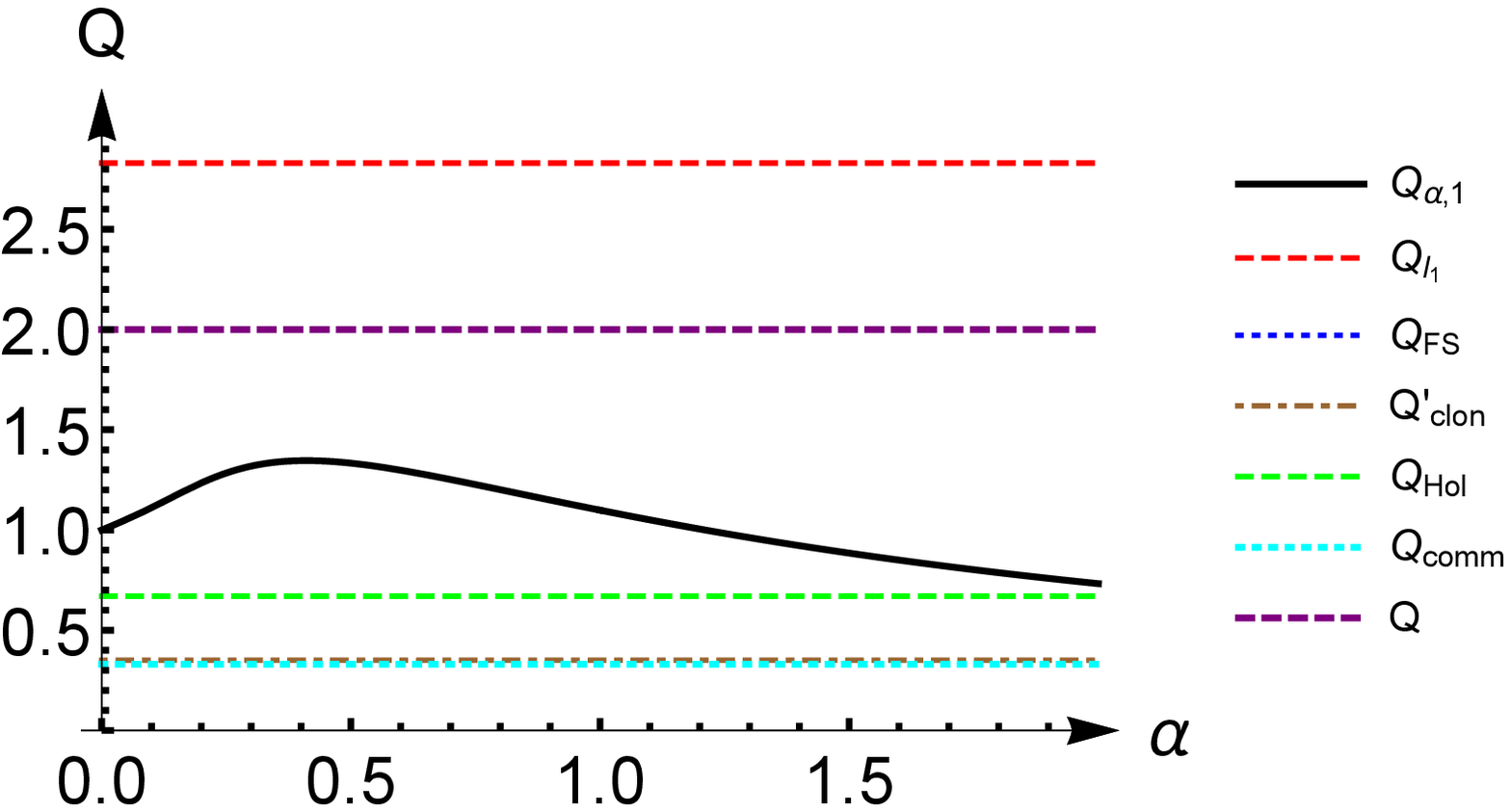}
%\caption{fig2}
\end{minipage}%
}%
\centering
\caption{ The quantumness of ensembles based on different quantifiers for a pure-state ensemble: (a) the B92 ensemble $\mathcal{E}_{\mathrm{B92}}$; (b) the diagonal ensemble $\mathcal{E}_{\mathrm{diag}}$; (c) the trine ensemble $\mathcal{E}_{\mathrm{trine}}$; (d) the BB84 ensemble $\mathcal{E}_{\mathrm{BB84}}$; (e) the tetrad ensemble $\mathcal{E}_{\mathrm{tetrad}}$; (f) the six-state ensemble $\mathcal{E}_{\mathrm{six}}$. The Q-axis denotes the quantumness with respect to various quantifiers.}
\end{figure}

\noindent {\bf 3. Conclusions}

Following the ideas in \cite{SY2021}, we have employed the
generalized $\alpha$-$z$-relative R\'enyi entropy of coherence of
Gram matrix to quantify the quantumness of pure-state ensembles and
explored its basic properties. Furthermore, we have calculated the
newly-defined quantumness for six ensembles and presented the
explicit formulas with parameter $\alpha$. These ensembles
arise in quantum cryptography or quantum measurement. We have
plotted the images of these quantumness measures as a function of
$\alpha$. It is found that for fixed $\alpha$,
$Q_{\alpha,1}(\mathcal{E}_{\mathrm{B92}})$ is always the minimum,
while $Q_{\alpha,1}(\mathcal{E}_{\mathrm{six}})$ is always the
maximum. There is an order of the corresponding quantities.
Moreover, we have also compared the quantumness of the six ensembles
with other quantumness quantifiers. It can be seen that different
quantifiers may yield different orderings of the quantumness for the
six ensembles. By plotting the images of the quantumness based on
various quantifiers for the same chosen pure-state ensemble
respectively, we have observed that the curves of $Q_{\alpha,1}(\cdot)$
intersects with the lines of other quantifiers at
different $\alpha$. This fact highlights the complexity and
subtlety of the quantumness measure since different quantifiers may
capture different aspects of the ensemble.

\textcolor{blue}{Our result enforces the previous finding in Refs.
\cite{BD1995,LGL2006} that though the density matrix of two
ensembles are identical, they differ in physics. In their work, they
used the general fluctuations, to distinguish them. It may be a
future topic to see if there is any link between the two quantities,
the fluctuation, and our quantumness defined via the generalized
$\alpha$-$z$-relative R\'enyi entropy of coherence.}

\textcolor{blue}{The quantumness of pure-state ensembles defined in
this paper may play a very important role in quantum information,
such as quantum key distribution \cite{BCH1984,KLC2021}, quantum
secure direct communication \cite{LGL2002,SYB2022}. It is more
capable than entanglement \cite{PJY2021,HWJ2020} in that they can
enable the secure transfer of information. It may also shed some
light on understanding the nature of measurement in quantum
mechanics \cite{LG2021,ZL2021}.}

Since a general ensemble consists of mixed quantum states, it is
necessary to extend our results from pure-state ensembles to the
case of mixed state ensembles. This important issue deserves further
study.

\noindent

%=============================================================================%
\subsubsection*{Acknowledgements}
The authors would like to express their sincere gratitude to the
anonymous referees for their valuable comments and suggestions,
which have greatly improved this paper. This work was supported by
National Natural Science Foundation of China (Grant Nos. 12161056,
11701259, 12075159, 12171044); Jiangxi Provincial Natural Science
Foundation (Grant No. 20202BAB201001); Beijing Natural Science
Foundation (Grant No. Z190005); Academy for Multidisciplinary
Studies, Capital Normal University; the Academician Innovation
Platform of Hainan Province; Shenzhen Institute for Quantum Science
and Engineering, Southern University of Science and Technology
(Grant No. SIQSE202001).

%=============================================================================%

%=============================================================================%
\subsubsection*{Competing interests}
The authors declare no competing interests.

%===========================================================================%

%=============================================================================%
\subsubsection*{Data availability}
Data sharing not applicable to this article as no datasets were
generated or analysed during the current study.

%===========================================================================%

\textcolor{blue}{
\appendix
\subsubsection* {\bf Appendix: Proof of the subadditivity of the quantumness $Q_{\alpha,z}(\cdot)$}}
\textcolor{blue}{According to Eq. (\ref{eq6}), we have
\begin{align*}
Q_{\alpha,z}^{'}(\mathcal{E}\otimes\mathcal{F})&=\underset{\sigma_1\in\mathcal{I}_1,\sigma_2\in\mathcal{I}_2}{\min}\frac{f^{\frac{1}{\alpha}}_{\alpha,z}(G_{\mathcal{E}\otimes
\mathcal{F}},\sigma_1\otimes \sigma_2)-1}{nm(\alpha-1)},\\
Q_{\alpha,z}^{'}(\mathcal{E})&=\underset{\sigma_1\in\mathcal{I}_1}{\min}\frac{f^{\frac{1}{\alpha}}_{\alpha,z}(G_\mathcal{E},
\sigma_1)-1}{n(\alpha-1)},\\
Q_{\alpha,z}^{'}(\mathcal{F})&=\underset{\sigma_2\in\mathcal{I}_2}{\min}\frac{f^{\frac{1}{\alpha}}_{\alpha,z}(G_\mathcal{F},
\sigma_2)-1}{m(\alpha-1)},
\end{align*}
where $\mathcal{I}_1$ and $\mathcal{I}_2$ denotes the set of
incoherent states on the $m$-dimensional and $n$-dimensional Hilbert
spaces, respectively. By the tensor multiplicability of the Gram
matrix, i.e.,
$G_{\mathcal{E}\otimes\mathcal{F}}=G_{\mathcal{E}}\otimes
G_{\mathcal{F}}$, we have
\begin{align*}
&f^{\frac{1}{\alpha}}_{\alpha,z}(G_{\mathcal{E}\otimes
\mathcal{F}},\sigma_1\otimes \sigma_2)\\
&=\{\mathrm{Tr}[(\sigma_1\otimes\sigma_2)^{\frac{1-\alpha}{2z}}{G_{\mathcal{E}\otimes
\mathcal{F}}}^{\frac{\alpha}{z}}(\sigma_1\otimes\sigma_2)^{\frac{1-\alpha}{2z}}]^{z}\}^{\frac{1}{\alpha}}\\
&=[\mathrm{Tr}({\sigma_1}^{\frac{1-\alpha}{2z}}{G_{\mathcal{E}}}^{\frac{\alpha}{z}}{\sigma_1}^{\frac{1-\alpha}{2z}})^{z}]^{\frac{1}{\alpha}}\cdot
[\mathrm{Tr}({\sigma_2}^{\frac{1-\alpha}{2z}}{G_{\mathcal{F}}}^{\frac{\alpha}{z}}{\sigma_2}^{\frac{1-\alpha}{2z}})^{z}]^{\frac{1}{\alpha}}\\
&=f_{\alpha,z}^{\frac{1}{\alpha}}(G_\mathcal{E},\sigma_1)\cdot
f_{\alpha,z}^{\frac{1}{\alpha}}(G_\mathcal{F},\sigma_2).
\end{align*}
So in order to prove the subadditivity, we only need to prove that
\begin{align}\label{eq15}
&\underset{\sigma_1\in\mathcal{I}_1,\sigma_2\in\mathcal{I}_2}{\min}\frac{f_{\alpha,z}^{\frac{1}{\alpha}}(G_\mathcal{E},\sigma_1)\cdot
f_{\alpha,z}^{\frac{1}{\alpha}}(G_\mathcal{F},\sigma_2)-1}{nm(\alpha-1)}\notag
\\
&\le\underset{\sigma_1\in\mathcal{I}_1}{\min}\frac{f_{\alpha,z}^{\frac{1}{\alpha}}(G_\mathcal{E},\sigma_1)-1}{n(\alpha-1)}
+\underset{\sigma_2\in\mathcal{I}_2}{\min}\frac{f_{\alpha,z}^{\frac{1}{\alpha}}(G_\mathcal{F},\sigma_2)-1}{m(\alpha-1)}.
\end{align}
Case (i): $0<\alpha<1,z>0$. Since the matrix
${\sigma}^{\frac{1-\alpha}{2z}}{\rho}^{\frac{\alpha}{z}}
{\sigma}^{\frac{1-\alpha}{2z}}$ has real, non-negative eigenvalues,
we obtain
$f_{\alpha,z}^{\frac{1}{\alpha}}(G_\mathcal{E},\sigma_1)\geq 0$ and
$f_{\alpha,z}^{\frac{1}{\alpha}}(G_\mathcal{F},\sigma_2)\geq 0$.
Noting that $f_{\alpha,z}^{\frac{1}{\alpha}}(\rho,\sigma)\leq 1$
when $0<\alpha<1$, we have $0\leq
f_{\alpha,z}^{\frac{1}{\alpha}}(G_\mathcal{E},\sigma_1)\leq 1$ and
$0\leq f_{\alpha,z}^{\frac{1}{\alpha}}(G_\mathcal{F},\sigma_2)\leq
1$, which implies that
\begin{equation}\label{eq16}
(f_{\alpha,z}^{\frac{1}{\alpha}}
(G_\mathcal{E},\sigma_1)-n)(f_{\alpha,z}^{\frac{1}{\alpha}}(G_\mathcal{F},\sigma_2)-m)\geq(1-n)(1-m)
\end{equation}
for each $\sigma_1\in\mathcal{I}_1$ and $\sigma_2\in\mathcal{I}_2$.
Hence, Eq. (\ref{eq15}) holds.
\\Case (ii): $1<\alpha\leq 2,z>0$. Since the completely mixed state $\sigma_*=I/d$ is a diagonal matrix, which is an incoherent state, we have
$\underset{\sigma\in\mathcal{I}}{\min}f_{\alpha,z}^{\frac{1}{\alpha}}(\rho,\sigma)\leq
f_{\alpha,z}^{\frac{1}{\alpha}}(\rho,\sigma_*)=
(d^{\alpha-1}{\mathrm{Tr}(\rho^{\alpha})})^{\frac{1}{\alpha}}\leq
d$. Noting that $ f_{\alpha,z}^{\frac{1}{\alpha}}(\rho,\sigma)\geq1$
when $\alpha>1$, we have
$1\leq\underset{\sigma_1\in\mathcal{I}_1}{\min}f_{\alpha,z}^{\frac{1}{\alpha}}(G_\mathcal{E},\sigma_1)\le
n$ and
$1\leq\underset{\sigma_2\in\mathcal{I}_2}{\min}f_{\alpha,z}^{\frac{1}{\alpha}}(G_\mathcal{F},\sigma_2)\le
m$, which implies that
\begin{align*}
&\underset{\sigma_1\in\mathcal{I}_1}{\min}f_{\alpha,z}^{\frac{1}{\alpha}}(G_\mathcal{E},\sigma_1)\cdot
\underset{\sigma_2\in\mathcal{I}_2}{\min}f_{\alpha,z}^{\frac{1}{\alpha}}(G_\mathcal{F},\sigma_2)-1\notag
\\
&\leq
m\left(\underset{\sigma_1\in\mathcal{I}_1}{\min}f_{\alpha,z}^{\frac{1}{\alpha}}(G_\mathcal{E},\sigma_1)-1\right)
+n\left(\underset{\sigma_2\in\mathcal{I}_2}{\min}f_{\alpha,z}^{\frac{1}{\alpha}}(G_\mathcal{F},\sigma_2)-1\right),
\end{align*}
and thus Eq. (\ref{eq15}) holds.}

\textcolor{blue}{In either case, we have proved Eq.(\ref{eq15}), and
so Eq.(\ref{eq7}) is established. This completes the proof. $\Box$}

\end{document}